\documentclass[prd,aps,twocolumn,nofootinbib,showpacs]{revtex4}

\usepackage{amsmath,graphicx,color,epsfig}

\begin{document}
\title{Setting the Renormalization Scale in pQCD:\\
Comparisons of the Principle of Maximum Conformality \\
with the Sequential Extended Brodsky-Lepage-Mackenzie Approach}

\author{Hong-Hao Ma}
\email{mahonghao@cqu.edu.cn}
\author{Xing-Gang Wu}
\email{wuxg@cqu.edu.cn}
\author{Yang Ma}
\email{mayangluon@cqu.edu.cn}

\address{Department of Physics, Chongqing University, Chongqing 401331, P. R. China}
\address{Institute of Theoretical Physics, Chongqing University, Chongqing 401331, P.R. China}

\author{Stanley J. Brodsky}
\email{sjbth@slac.stanford.edu}
\affiliation{SLAC National Accelerator Laboratory, Stanford University, Stanford, California 94039, USA}

\author{Matin Mojaza}
\email{mojaza@nordita.org}
\affiliation{Nordita, KTH Royal Institute of Technology and Stockholm University,
Roslagstullsbacken 23, SE-10691 Stockholm, Sweden}

\date{\today}

\begin{abstract}

A key problem in making precise perturbative QCD (pQCD) predictions is how to set the renormalization scale of the running coupling unambiguously at each finite order. The elimination of the uncertainty in setting the renormalization scale in pQCD will greatly increase the precision of collider tests of the Standard Model and the sensitivity to new phenomena. Renormalization group invariance requires that predictions for observables must also be independent on the choice of the renormalization scheme. The well-known Brodsky-Lepage-Mackenzie (BLM) approach cannot be easily extended beyond next-to-next-to-leading order of pQCD. Several suggestions have been proposed to extend the BLM approach to all-orders. In this paper we discuss two distinct methods. One is based on the ``Principle of Maximum Conformality'' (PMC), which provides a systematic all-orders method to eliminate the scale- and scheme- ambiguities of pQCD. The PMC extends the BLM procedure to all orders using renormalization group methods; as an outcome, it significantly improves the pQCD convergence by eliminating renormalon divergences. An alternative method is the ``sequential extended BLM" (seBLM) approach, which has been primarily designed to improve the convergence of pQCD series. The seBLM, as originally proposed, introduces auxiliary fields and follows the pattern of the $\beta_0$-expansion to fix the renormalization scale. However, the seBLM requires a recomputation of pQCD amplitudes including the auxiliary fields; due to the limited availability of calculations using these auxiliary fields, the seBLM has only been applied to a few processes at low-orders. In order to avoid the complications of adding extra fields, we propose a modified version of seBLM which allows us to apply this method to higher-orders. We then perform detailed numerical comparisons of the two alternative scale-setting approaches by investigating their predictions for the annihilation cross section ratio $R_{e^+e^-}$ at four-loop order in pQCD.

\pacs{12.38.Bx, 12.38.Aw, 11.15.Bt}

\end{abstract}

\maketitle

\tableofcontents

\clearpage

\section{Introduction}

Renormalization group (RG) invariance is a central principle of quantum field theories: there cannot be any renormalization scheme or renormalization scale ambiguity in the predictions for physical observables. However, this principle is typically violated if one uses conventional scale-setting methods in QCD due to the unavoidable truncation of the pQCD series. Thus, a key problem in making precise pQCD predictions is how to set the renormalization scale of the running coupling properly at each perturbative order without introducing any unphysical dependence on the choice of renormalization scheme. The elimination of the uncertainty in setting the renormalization scale in pQCD will greatly increase the precision of collider tests of the Standard Model and the sensitivity to new phenomena. A review of the QCD renormalization scale-setting problem can be found in Ref.~\cite{pmcreview}.

One of the earliest approaches to solve the scale-setting problem in pQCD is known today as the BLM approach, suggested by Brodsky, Lepage and Mackenzie in Ref.~\cite{BLM}. The BLM approach was inspired by its counterpart in Abelian quantum electrodynamics (QED) -- the Gell Mann-Low scheme~\cite{gml}. In QED, only vacuum-polarization insertions contribute to coupling constant renormalization. In the BLM work, the optimal scales for the running coupling in non-Abelian theory are set at each order in analogy to QED by absorbing all of the contributions of the $\beta$-function of the process into the running coupling. In the BLM procedure, the $n_f$ contributions of fermion loops were used to identify the leading $\{\beta_i\}$-terms. However this identification is only effective up to next-to-leading order (NLO). The BLM method was later extended to next-to-next-to-leading order (NNLO) by Brodsky and Lu~\cite{scale}, who also observed that the BLM predictions are independent of the choice of the scheme if one relates observables to each other via commensurate scale relations. However, this method cannot be unambiguously extended to pQCD series at very high orders, and thus an extra procedure to distinguish the $\{\beta_i\}$-terms in a pQCD series is required.

A proposed extension of BLM is a method called ``sequential extended BLM" (seBLM)~\cite{seBLM1}. The purpose of seBLM is to optimize the pQCD convergence. In order to accomplish this, a ``large $\beta_0$-approximation"~\cite{approximation1,approximation2}, with slight modifications, is adopted to deal with the pQCD series. [Here $\beta_0$ denotes the first perturbative coefficient of the running coupling $\beta$-function.] However, the seBLM does not distinguish whether the $n_f$-terms in a pQCD expansion are in fact related to the renormalization group (RG) $\beta$-function, and thus all $n_f$-terms are rearranged into $\{\beta_i\}$-terms. Subsequently the $\{\beta_i\}$-terms are resummed into the running coupling following the scheme of the $\beta_0$-expansion. Due to ambiguities in determining the coefficients of the $\{\beta_i\}$-terms at higher-orders, new colored fields are introduced to fix the seBLM scales. A detailed review of the seBLM procedures will be given in this paper; we will also point out several shortcomings of the seBLM method, which constrains its current applicability.

If one can unambiguously identify all of the $\{\beta_i\}$-terms in a pQCD expansion, then they can be systematically eliminated by shifting and thus setting the renormalization scale at each order in pQCD. The remaining pQCD series will then match the terms of the corresponding ``conformal" series with $\beta=0$. This is the main idea behind the ``Principle of Maximum Conformality (PMC)"~\cite{pmc1,pmc2,pmc3,pmc4,pmc5,BMW,BMW2}. The PMC explains why the BLM method works so successfully at the NLO level, and it can be applied for any high-energy processes up to any order~\cite{BMW}. For some recent N$^3$LO and N$^4$LO examples see Refs.~\cite{app1,app2,app3,app4,app5}.

When one applies PMC scale-setting, the scales of the running coupling in the pQCD series are shifted such that all contributions related to the $\beta$-function are resummed and only the coefficients which are RG invariant remain. These coefficients are called the ``conformal series" since they are identical to the coefficients obtained if the $\beta$ function were zero. The resulting PMC predictions are thus scheme-independent -- predictions under different schemes only differ by the appropriate shift of the renormalization scale. Thus the PMC numerical prediction is insensitive to the scheme choice. The PMC predictions also have the property that any residual scale dependence on the choice of (initial) renormalization scale is highly suppressed, even for low-order predictions. The scheme-independence of the PMC predictions is also confirmed by commensurate scale relations. An all-order demonstration of commensurate scale relations in pQCD can be found in Refs.~\cite{pmcreview,BMW2}. The PMC thus obeys standard RG-invariance and satisfies all RG-properties~\cite{pmccolloquium}. Furthermore, the pQCD convergence of pQCD series is improved due to the elimination of divergent $(n! \beta^n_i \alpha_s^n)$-renormalon terms.

Two ways to identify the nonconformal $\{\beta_i\}$-terms and to implement the PMC have been suggested. One method is based on the PMC-BLM correspondence (PMC-I)~\cite{pmc1,pmc4}, and the other more recent method is a theoretical improvement based on the $R_\delta$-scheme (PMC-II)~\cite{BMW,BMW2}. Both of these methods satisfy RG-invariance, but they resum the pQCD series in different steps. We have shown that PMC-I and PMC-II are numerically equivalent by comparing several high-energy processes up to the four-loop level~\cite{pmc6,app4}. A systematic demonstration of this equivalence will appear soon~\cite{equivalence}. In the following, we will adopt PMC-II for the discussions and simply call it PMC.

Recently, an approximate comparison of PMC and seBLM was published by Kataev and Mikhailov~\cite{Kataev1}; they concluded that the PMC method is ``questionable". However, we have found that their conclusions were based on a misunderstanding and a misuse of the PMC. It is clearly important to clarify these issues so that any misuse of the PMC method will be avoided. In this paper, we will analyze $R_{e^+ e^-}$ up to four-loop level, making a detailed comparison of the PMC and seBLM predictions. Since the seBLM is currently not applicable at the N$^3$LO level and beyond, we will suggest a new method for extending it to higher orders, based on lessons from the PMC. We will refer to this modified seBLM as `MseBLM'.

The remaining parts of this paper are organized as follows: In Sec. II, we will present a mini-review of the BLM-like scale-setting methods, BLM, PMC and \mbox{seBLM}. The main features of these BLM-like methods will be discussed, and an overview of the differences between the PMC and seBLM procedures will then be presented. We will show that the PMC and seBLM methods have some common features; in particular, the PMC and the seBLM $\beta$-patterns are exactly the same. We will also present \mbox{MseBLM} in this section. In Sec. III, we will present numerical results for $R_{e^+ e^-}$ up to the four-loop level to provide a detailed comparison of the PMC, seBLM and MseBLM predictions. Sec. IV is reserved for a summary and conclusions. Two appendices provide further computational details for seBLM.

\section{BLM-like scale settings}

Consider the $n_{\rm th}$-order pQCD approximant $\varrho_n$ for a typical physical observable $\varrho$,
\begin{equation} \label{iexpansion}
\varrho_n=r_0 a^{p}(\mu)+\sum_{i=1}^{n} r_i(\mu) a^{p+i}(\mu),
\end{equation}
where $a=\alpha_s/4\pi$ and $\mu$ stands for the initial choice of scale. The coefficient $r_0$ is the tree-level result, $p$ is the power of the coupling associated with the tree-level term, and $r_i(\mu)$ is the coefficient of the $i_{\rm th}$-loop correction.

In conventional scale-setting treatments, the renormalization scale is fixed to an initial guessed value which is usually taken as a typical momentum flow of the process. If one uses this method, the prediction is scheme dependent. Furthermore, the scheme and scale dependence of the coefficient $r_i(\mu)$ and the coupling constant do not exactly cancel at any fixed-order, leading to well-known ambiguities. In contrast, the BLM, PMC and seBLM scale-setting methods improve the pQCD predictions by eliminating such scheme and scale ambiguities.

\subsection{Basic arguments of BLM}

The renormalization scale for the running coupling is unambiguously set in QED by summing all vacuum polarization contributions, both proper and improper, into the photon propagator. Thus $\alpha(t) ={ \alpha(0) \over 1-\Pi (t)}$ in the Gell Mann-Low scheme~\cite{gml}, where $t$ is the virtuality of the exchanged photon and $\Pi(t)$ sums all $n_f$ fermion loop contributions. The running of the QED effective coupling is due to vacuum polarization alone -- only vacuum-polarization insertions contribute to the effective coupling~\cite{gml}. Following this observation, Brodsky, Lepage and Mackenzie pointed out that the $n_f$-dependence of the pQCD series can also be used at low orders as a guide to identify the $\beta_0$ and $\beta_1$ terms and thus set the scale of the pQCD prediction up to N$^2$LO~\cite{BLM}. This simple and straightforward $n_f$ method can be applied to processes that do not involve the three- or four-gluon couplings at leading order. The BLM method also ensures that the pQCD predictions analytically continue at $N_c \to 0$ correctly to Abelian theory~\cite{Brodsky:1997jk}.

We will take the NLO pQCD prediction for a typical single-scale physical observable $\varrho_1$ to illustrate the BLM procedure. At NLO level, the pQCD prediction can be re-expanded as
\begin{eqnarray}
\varrho_1 &=& r_{0}a^{p}(\mu) \left[1+\left(A n_{f}+B\right) a(\mu) \right].
\end{eqnarray}
The $n_f$-term is due to light quark vacuum-polarization insertions. Any initial scale choice and any renormalization scheme, including dimensional regularization, can be used for the prediction. The coefficients $A$ and $B$ are in general different under different schemes; however, the BLM prediction ${\varrho}_1$ is unchanged for any choice of scheme due to commensurate scale relations~\cite{scale}.

All of the $n_f$-terms can be resummed into the running coupling with the help of the one-loop $\alpha_s$-running coupling
\begin{equation}
a(\mu_{\rm BLM})=\frac{a(\mu)} {1+ {\beta_0} a(\mu) \ln \left({\mu^{2}_{\rm BLM}}/{\mu^2}\right)} , \label{resum}
\end{equation}
where $\beta_0=11-2n_f/3$. One then obtains
\begin{equation}
\varrho_1 = r_{0}a^{p}(\mu_{\rm BLM}) \left[1+ r^*_{1} {a(\mu_{\rm BLM})} \right],
\end{equation}
where $\mu_{\rm BLM}=\mu\exp({3A}/{p})$ and $r_1^*=\frac{33}{2}A +B$. The BLM scale $\mu_{\rm BLM}$ is thus determined solely by $A$. The term $33A/2$ in $r_1^*$ serves to remove those contributions which renormalize the running coupling, and the resulting $a(\mu_{\rm BLM})$ is the predicted value of the running coupling. Eq.(\ref{resum}) indicates that the BLM coupling $a(\mu_{\rm BLM})$, and thus the BLM prediction $\varrho_1$, is independent of the initial scale $\mu$, as is readily checked.

This approach of using the $n_f$-terms as a guide to resum the series through the RG-equation of $\alpha_s$ cannot be unambiguously extended to higher orders. One reason is that the $n_f$-series and the $\{\beta_i\}$-series are not {\it a priori} one-to-one. Another issue is that $n_f$-terms appear at higher orders from loops which are ultraviolet finite but are not associated with the $\beta$-function of the running coupling. Thus reactions with multi-gluon couplings are more difficult to analyze using BLM because quark loops appear in high-order corrections to the multi-gluon vertex as well as in the propagator insertions~\cite{binger}. Scale setting for the BFKL Pomeron intercept provides such an example~\cite{pomeron1,pomeron2}.

Thus, it is necessary to modify BLM at higher-orders. We will discuss two suggestions, PMC and seBLM. We shall first present an overview of those two suggestions, and then present the PMC and seBLM features and their consequences sequentially. In this discussion we will ignore quark mass terms and their renormalization.

\subsection{An overview of PMC and seBLM} \label{SecII.B}

The purposes of PMC and seBLM are different. The PMC is designed to solve the renormalization scheme- and scale- ambiguities, whereas the seBLM is designed to improve the pQCD convergence. Both the PMC and seBLM utilize the $\{\beta_i\}$-series to achieve these goals, rather than the simpler $n_f$-terms. In the case of processes where the $\{\beta_i\}$-terms for the quark anomalous dimension and the QCD $\beta$-function are entangled with each other, extra steps have to be taken to distinguish those $\{\beta_i\}$-terms~\cite{app1}.

When one applies the PMC or seBLM, two steps are needed to fix the renormalization scale. The first step in both cases is to fix the $\beta$-pattern at each perturbative order and determine the coefficients of all $\{\beta_i\}$-terms in the $\beta$-pattern. However, PMC and seBLM use quite different methods to accomplish this step:
\begin{itemize}
\item The PMC observes that the $\beta$-pattern at each order originates from a specific pattern and superposition of the $\{\beta_i\}$-terms coming from all the lower-order $\alpha_s$-factors, due to its running behavior. The running behaviors are governed by the fundamental RG-equation, and one can thus identify the $\beta$-pattern up to all orders without any ambiguities. As we shall discuss below, this procedure can be carried out systematically by generalizing the definition of $\overline{\rm MS}$ dimensional regularization to include extra subtraction terms $\delta_{j=1,2,\cdots}$. The coefficients of $\delta^{m=1,2,\cdots}_{j=1,2,\cdots}$ then isolate the $\{\beta_i\}$-terms to a particular $\alpha_s$-order~\cite{BMW,BMW2}.

   The coefficients of the $\{\beta_i\}$-terms at each order can also be determined from the $n_f$-power series at the same order, which are calculated under a certain renormalization scheme, such as the $\overline{\rm MS}$-scheme~\cite{MSbar}. The expressions for the $\beta_{0,1,2,3}$ as a function of $n_f$ in the $\overline{\rm MS}$-scheme can be found in Refs.\cite{beta0,beta00,beta01,beta02,beta1,beta11,beta2,beta3,beta4}. There is a subtlety regarding $n_f$-terms that are unrelated to the $\alpha_s$-renormalization, which must be kept unchanged when applying the PMC. Special degeneracy relations among different order terms ensure the exact one-to-one correspondence between the $n_f$-terms and the $\{\beta_i\}$-terms at the same order, so that all $\{\beta_i\}$-coefficients can be unambiguously fixed up to all orders~\cite{BMW}. It has been demonstrated that the degeneracy relations are not specific to dimensional regularization schemes, but are general features of perturbation theory~\cite{BMW2}.

\item The seBLM fixes the $\beta$-pattern by identifying the equivalent $\beta_0$-powers of the $\{\beta_i\}$-terms via the relation $\beta_i\sim\beta^{i+1}_0$. All of the possible $\{\beta_i\}$-terms, whose equivalent $\beta_0$-powers are equal or less than the maximum $\beta_0$-power of the considered order, form the $\beta$-pattern at each order. In fact the seBLM $\beta$-pattern is the same as the PMC one. According to the RG-equation, the running coupling at different scales are related by the following displacement equation:
\begin{widetext}
\begin{eqnarray}
a(\mu_2) &=& a(\mu_1)- \beta_{0} \ln\left(\frac{\mu_2^{2}} {\mu_1^2}\right) a^{2}(\mu_1) + \left[\beta^2_{0} \ln^2 \left(\frac{\mu_2^{2}}{\mu_1^2}\right) -\beta_{1} \ln\left(\frac{\mu_2^{2}} {\mu_1^2}\right) \right] a^{3}(\mu_1) + \nonumber\\
&& \left[-\beta^3_{0} \ln^3 \left(\frac{\mu_2^{2}} {\mu_1^2}\right) +\frac{5}{2} \beta_{0}\beta_{1} \ln^2\left(\frac{\mu_2^{2}} {\mu_1^2}\right) -\beta_{2} \ln\left(\frac{\mu_2^{2}}{\mu_1^2}\right)\right] a^{4}(\mu_1) +{\cal O}(a^{5}) , \label{scaledis}
\end{eqnarray}
\end{widetext}
    where $\mu_1$ and $\mu_2$ are two arbitrary scales. At each order, the equivalent $\beta_0$-powers are the same for all $\{\beta_i\}$-terms. For example, at order $a^4$, $\beta_2 \sim \beta_0\beta_1 \sim \beta^3_0$. Because the maximum $\beta_0$-power at each order is the same for PMC and seBLM, the superposition of the $\alpha_s$-displacement from all lower orders, which is used in PMC, will, as we will see, result in the $\beta$-pattern of seBLM. From this point of view, the PMC provides the underlying principle for the seBLM $\beta$-pattern.

    The seBLM coefficients of the $\{\beta_i\}$-terms are also determined from the known $n_f$-power series, but in a quite different way than the PMC: At the NLO level, only the $\beta_0$-term needs to be determined, and it can be directly fixed by the $n^{1}_f$-term. At the NNLO level, the $\beta^2_0$-term can be fixed by the $n^{2}_f$-term, but the $\beta_1$- and $\beta_0$-terms cannot be unambiguously fixed by the $n^{1}_f$-term alone, since both $\beta_0$ and $\beta_1$ are linear functions of $n^{1}_f$. To solve this problem, the seBLM method introduces $\tilde{n}_g$ multiplets of fermions in the adjoint representation of the color-group, resembling the gluino of supersymmetric Yang-Mills theory (which makes part of the minimal supersymmetric standard model, MSSM). The coefficients for the $\beta_1$- and $\beta_0$- terms are thus fixed by recalculating pQCD series with $\tilde{n}_g$ gluinos and by using the resulting $n^{1}_f$- and $\tilde{n}^{1}_g$- terms. The gluinos are introduced as a technical device to fix the $\{\beta_i\}$-coefficients; however this procedure introduces process-dependent complex calculations into higher-order QCD corrections. Thus in order to apply the seBLM, the pQCD corrections need to be recomputed with the new fields in order to extract the $\tilde{n}_g$-dependence. At present, the $\tilde{n}_g$-dependent $\beta_{i}$-function is known to three-loops; {\it i.e.}, up to $\beta_2$, and only the Adler $D$-function has been calculated with $\tilde{n}_g$-dependence up to NNLO level. Thus seBLM can only be applied at this time up to NNLO for $D$-function-derived processes, such as $R_{e^+ e^-}$ and the Bjorken polarized sum rule~\cite{Kataev1}.

\end{itemize}

For the second step, one needs to find the correct way to resum the relevant $\{\beta_i\}$-terms, determined from the first step, into the running coupling. The PMC and the seBLM take quite different paths:
\begin{itemize}
\item The PMC notes that only those $\{\beta_i\}$-terms which are related with the renormalization of the running coupling should be absorbed into the running coupling. They are eliminated through the RG-equation, and the resultant PMC scales are functions of the running coupling and are in general different for different orders. One can choose any initial renormalization scale to do the pQCD calculation as long as its value is large enough to ensure pQCD applicability. The final resummed result, however, has no or very small residual dependence on the choice of the initial renormalization scales.

\item The seBLM treats all the $\{\beta_i\}$-terms on equal footing without distinguishing whether they should be incorporated into the renormalization of the running coupling. The seBLM scales are obtained by adopting the ``large $\beta_0$-approximation"~\cite{approximation1,approximation2}; i.e. using the equivalent $\beta_0$-powers for the $\{\beta_i\}$-terms as a guide. The highest $\beta_0$-powers are eliminated first, then the one-order-lower $\beta_0$-terms, and so on. Thus the seBLM scales are effective scales, and one cannot {\it a priori} decide whether they will lead to the optimal behavior of the running coupling. In fact, one finds that the seBLM convergence is not as expected, even for the quantity $R_{e^+ e^-}$. Thus an extra treatment, called $x$-BLM, to further improve the seBLM pQCD convergence was suggested in the original seBLM paper~\cite{seBLM1}.
\end{itemize}

\subsection{PMC scale setting}

PMC scale-setting provides the underlying principle for BLM, a procedure that can be unambiguously applied to any order~\cite{BMW,BMW2,pmcreview}. The PMC utilizes the identified $\{\beta_i\}$-terms and the RG-equation to determine the value of the running coupling at each order and thus the ``physical" scales of the process. At the NLO level, one only requires the $\beta_0$-term; absorbing the $\beta_0$-term into the scale of the running coupling is equivalent to eliminate the $n^1_f$-term, which explains why BLM works so well at the NLO level.

A key step is to identify the $\beta$-terms in the pQCD prediction, thus distinguishing the ``nonconformal" versus the ``conformal" $\beta$-independent terms. To do this, the PMC introduces a generalized dimensional renormalization scheme, the $R_\delta$-scheme, where an arbitrary constant $-\delta$ is subtracted in addition to the standard subtraction $(\ln 4\pi - \gamma_{E})$ used in the $\overline{\rm MS}$-scheme. The $\delta$-subtraction defines an infinite set of MS-like schemes. The PMC scales for different $R_\delta$-schemes, e.g. $R_{\delta_1}$- and $R_{\delta_2}$-schemes, differ only by a factor $e^{(\delta_1-\delta_2)/2}$. This scale relation ensures the scheme independence of pQCD predictions among different schemes. Moreover, the scale displacement between couplings in any $R_\delta$-scheme reveals all the $\{\beta_i\}$-terms pertaining to a specific order~\cite{BMW,BMW2}. By collecting all of the $\{\beta_i\}$-terms that occur at a given order, one can identify the $\beta$-pattern of the RG, and thus compute each PMC scale order by order,

According to PMC scale-setting, the explicit $\beta$-pattern at each order for the pQCD prediction of the observable $\varrho$ can be written as
\begin{widetext}
\begin{eqnarray}
\varrho_n &= &r_{0,0} + r_{1,0} a(\mu) + \left[r_{2,0} + \beta_0 r_{2,1} \right] a^2(\mu) + \left[r_{3,0} + \beta_1 r_{2,1} + 2 \beta_0 r_{3,1} + \beta _0^2 r_{3,2} \right] a^3(\mu) \nonumber\\
&& +\left[r_{4,0} + \beta_2 r_{2,1} + 2\beta_1 r_{3,1} + \frac{5}{2} \beta_1 \beta_0 r_{3,2} +3\beta_0 r_{4,1}+ 3 \beta_0^2 r_{4,2} + \beta_0^3 r_{4,3} \right] a^4(\mu) + \cdots \label{PMCbetapattern}
\end{eqnarray}
\end{widetext}
where the $r_{i,0}$ are the conformal coefficients, while $r_{i,j\neq0}$ are the ones related to the running coupling renormalization. The degeneracy relations among the non-conformal coefficients at different orders are implicitly adopted. The $\beta$-pattern fixed by Eq.(\ref{PMCbetapattern}) at each order is dictated by the RG-equation. It is thus natural to call such a $\beta$-pattern the RG-$\beta$-pattern. It is noted that the RG-$\beta$-pattern is determined from a superposition of displacements in the running couplings at each order; thus only those $\{\beta_i\}$-terms which determine the correct running behavior of the coupling are kept. There are cases where the coefficients $r_{i,j}$ are exactly zero for the $\{\beta_i\}$-terms; i.e. there are ``missing" $\{\beta_i\}$-terms in specific processes. This only indicates that those terms have no contributions to the renormalization of the running coupling. This explains why in Ref.\cite{Kataev1}, what the authors refer to as the``correct PMC", is actually an invalid procedure. Since the authors~\cite{Kataev1} transform all $n_f$-terms into $\{\beta_i\}$-terms which brings unrelated $\{\beta_i\}$-terms into the RG-$\beta$-pattern, thus explicitly breaking the PMC procedure.

Using PMC scale setting, which follows the pattern dictated by the RG, all lower-order running couplings can be resummed into effective couplings $a^k(Q_k)$ as
\begin{widetext}
\begin{eqnarray}
r_{k,0} a^k(\mu) + r_{k,0}\ k \ a^{k-1}(\mu) \beta(a) \left \{ R_{k,1} +\Delta_k^{(1)}(a) R_{k,2} + \cdots + \Delta_k^{(n-1)}(a) R_{k,n} \right \} &=& r_{k,0} a^{k}(Q_k) \ ,
\label{korder}
\end{eqnarray}
where $\beta(a) = - a^2(\mu) \sum_{i=0}^\infty \beta_i a^{i}(\mu)$ and $Q_k$ is the PMC scale at $a^k$-order, given by
\begin{eqnarray}
\label{exactscales}
\ln \frac{Q_{k}^2}{\mu^2} &=& \frac{R_{k,1} + \Delta_k^{(1)}(a) R_{k,2}+\Delta_k^{(2)}(a) R_{k,3}+\cdots } {1+ \Delta_k^{(1)}(a) R_{k,1} + \left({\Delta_k^{(1)}(a)}\right)^2 (R_{k,2} -R_{k,1}^2) + \Delta_k^{(2)}(a)R_{k,1}^2 +\cdots } \ , \label{PMCscale} \\
R_{k,j} &=& (-1)^{j}\frac{r_{k+j, j}}{r_{k,0}} \ , \\
\Delta_k^{(1)}(a) &=& \frac{1}{2!} \left [ \frac{\partial \beta}{\partial a} + (k-1) \frac{\beta}{a}\right],\ \Delta_k^{(2)}(a) = \frac{1}{3!}\left [ \beta \frac{\partial^2 \beta}{\partial a^2} +Ê \left( \frac{\partial\beta}{\partial a} \right )^2 + 3(k-1) \frac{\beta}{a}\frac{\partial\beta}{\partial a} +(k-1)(k-2) \frac{\beta^2}{a^2}\right ] \ \cdots \ .
\end{eqnarray}
\end{widetext}
This shows that the PMC scales are in general different at different orders, as in QED. This is clear since the $\beta$-terms which control the behavior of the running coupling and the physical scales which set the virtuality of the propagators, as well as the number of effective flavors $n_f$, are usually different at each order.

The final pQCD PMC prediction for $\varrho$ then reads
\begin{eqnarray}
\varrho_n &=& r_{0,0} + r_{1,0} a(Q_1) + r_{2,0} a^2(Q_2) \nonumber\\
&&
+r_{3,0} a^3(Q_3) + r_{4,0}a^4(Q_4)+ \cdots \ . \label{PMCpQCD}
\end{eqnarray}
At the four-loop level, $Q_4$ remains unknown, since we need to know the five-loop coefficient $r_{5,1}$ to fix its value. This leads to some minor residual scale dependence, suppressed by the highest power in $a$. A practical choice of $Q_4$ is $Q_3$. As shown by Eq.(\ref{PMCscale}), the PMC scales themselves are perturbative series, which also introduce residual scale-dependence; however, this dependence is highly suppressed, even for lower-order analyses. Since the nonconformal contributions are absorbed into the scales, the PMC predictions have optimal pQCD convergence due to the elimination of divergent renormalon terms. More details and the properties of PMC may be found in Refs.\cite{pmc1,pmc2,pmc3,pmc4,pmc5,BMW,BMW2} and in a recent short review in Ref.\cite{pmc6}.

\subsection{seBLM and its modified version MseBLM}

As mentioned in the introduction, seBLM can only be applied at present to Adler $D$-function-derived processes up to three-loops~\cite{Kataev1,Kataev2,Kataev3,Kataev4}. In this subsection, we will take the three-loop pQCD prediction of a physical observable $\varrho$ to illustrate the seBLM procedure. Following Eq.(\ref{iexpansion}), the pQCD prediction $\varrho_n$ for $\varrho$ up to three-loop level can be written as
\begin{widetext}
\begin{eqnarray}
\varrho_n &=& r_0+r_1 \left(a(\mu) + r_2 a^{2}(\mu)+ r_3 a^{3}(\mu)+\cdots \right) \label{seBLMstart} \nonumber \\
&=& r_0+r_1 \left(a(\mu) + \left(\beta_0 \cdot r_{2}[1]+ r_{2}[0]\right) a^{2}(\mu) + \left(\beta_0^2\cdot r_{3}[2]+\beta_1\cdot r_{3}[0, 1]+ \beta_0\cdot r_{3}[1]+ r_{3}[0]\right) a^{3}(\mu) + \cdots \right), \label{seBLMpattern}
\end{eqnarray}
\end{widetext}
where $r_0$ is free of strong interactions, and for convenience, an overall factor $r_1$ is factored out of the pQCD corrections. The $\beta$-pattern at each order in the second line is suggested in Ref.~\cite{seBLM1}. The first argument $n_0$ of the coefficients $r_{n}[n_0,n_1,\cdots]$ correspond to the $\beta_0$-power, whereas the second one $n_1$ corresponds to the $\beta_1$-power, etc. We have omitted the arguments of the $\{\beta_i\}$-coefficients for brevity. In order to shorten the notation even further: if all the arguments of the coefficient $r_{n}[\cdots,m,0,\cdots, 0]$ to the right of the index $m$ are equal to zero, then we will omit those zero arguments for simplicity and write instead simply $r_{n}[\cdots,m]$. The coefficient $r_{n}[n_0,n_1,\cdots]$ is usually scale-dependent through terms dependent on $\ln Q^2/\mu^2$, where $Q$ stands for the typical momentum flow of the process.

The seBLM $\beta$-pattern, shown in Eq.(\ref{seBLMpattern}), is fixed by using the relation, $\beta_i\sim\beta^{i+1}_0$. By counting the equivalent $\beta_0$-powers for all the possible $\{\beta_i\}$-terms, the $\beta$-pattern is determined by following the decrement of the equivalent $\beta_0$-powers. For example: At the N$^2$LO level, we have only $\beta^1_0$; at the N$^3$LO level, we have $\beta^2_0$, $\beta_1\sim\beta^2_0$ and $\beta_0$; at the N$^4$LO level, we have $\beta^3_0$, $\beta_2\sim\beta^3_0$, $\beta_1\beta_0\sim\beta^3_0$, $\beta^2_0$, $\beta_1\sim\beta^2_0$, $\beta_0$; at the N$^5$LO level, we have $\beta^4_0$, $\beta_3\sim\beta^4_0$, $\beta_2\beta_0\sim\beta^4_0$, $\beta^2_1\sim\beta^4_0$, $\beta_1\beta^2_0\sim\beta^4_0$, $\beta^3_0$, $\beta_2\sim\beta^3_0$, $\beta_1\beta_0\sim\beta^3_0$, $\beta^2_0$, $\beta_1\sim\beta^2_0$, $\beta_0$; etc.

\begin{figure}[htb]
\centering
\includegraphics[width=0.48 \textwidth]{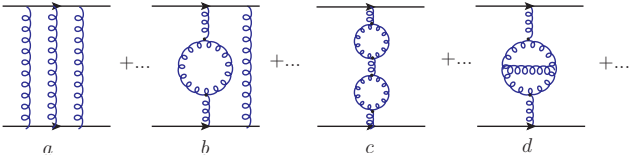}
\caption{Diagrammatic illustration for the $\beta$-pattern of the $\alpha_s^3$-correction contributing to the photon polarization operator $\Pi$~\cite{seBLM1}, following the relation $\beta_i\sim\beta_0^{i+1}$: (a) diagrams without any intrinsic renormalization contributions to $r_3[0]$; (b) diagrams contributing to the $\beta_0$-term $r_3[1]$; (c) diagrams contributing to the $\beta^2_0$-term $r_3[2]$; (d) diagrams contributing to the $\beta_1$-term $r_3[0,1]$. The ellipses denote other diagrams with the gluon/vertex renormalization. } \label{beta1}
\end{figure}

A diagrammatic illustration of the seBLM $\beta$-pattern can be motivated by the $\alpha_s$-corrections to the photon polarization operator $\Pi$~\cite{seBLM1}. It is based on the fact that the $\beta_i$-functions emerge together with certain $\alpha_s$-powers; {\it i.e.}, $(\beta_0 a)$, $(\beta_1 a^2)$, $\cdots$, $(\beta_i a^{i+1})$; at a fixed $\alpha_s$-order, all possible combinations of $\beta_i$-functions can thus be enumerated. Fig.(\ref{beta1}) shows the $\alpha^3_s$-corrections to the gluon polarization operator $\Pi$, in which the diagrams for the coefficients $r_3[0]$, $r_3[1]$, $r_3[2]$ and $r_3[0,1]$ are presented. Another explanation of the seBLM $\beta$-pattern can be based on an analysis of the Casimir structures of the Adler $D$-function; see Ref.\cite{Kataev1}. Those two explanations for the seBLM $\beta$-pattern are illuminating; however, they cannot explain why such $\beta$-pattern would be applicable to all high-energy processes. Note that the seBLM $\beta$-pattern is the same as the PMC RG-$\beta$-pattern. As has been discussed in Sec.\ref{SecII.B}, the PMC provides the underlying reason for the seBLM $\beta$-pattern.

\subsubsection{Determination of seBLM $\{\beta_i\}$-coefficients}

In practice, the seBLM $\{\beta_i\}$-coefficients $r_{n}[n_0,n_1,\cdots]$ are fixed by using the full $n_f$-power series at the same order without distinguishing whether those $n_f$-terms pertain to the RG $\beta$-function. As we have noted, this is the underlying reason why the seBLM differs from the PMC.

At the two-loop level, the $\beta_0$-coefficient $r_{2}[1]$ can be fixed by the linear $n_f$-term $c_{2,1} n_f$ in the two-loop $n_f$-power series
\begin{eqnarray}
r_{2}= c_{2,1} n_f + c_{2,0} \ , \label{r2nf}
\end{eqnarray}
where the expansion coefficients $c_{2,0}$ and $c_{2,1}$ for the case of Adler $D$-function can be found in Ref.\cite{Dfun}.

At the three-loop level, the $\beta_0^2$-coefficient $r_{3}[2]$ can be fixed by the squared $n_f^2$-term $c_{3,2} n^2_f$; however, the $\beta_0$ and $\beta_1$ coefficients $r_{3}[1]$ and $r_{3}[0,1]$ cannot be unambiguously fixed by a single linear $n_f$-term $c_{3,1} n_f$ in the third-order $n_f$-power series
\begin{eqnarray}
r_{3}= c_{3,2} n^2_f + c_{3,1} n_f + c_{3,0}, \label{r3nf}
\end{eqnarray}
where the expansion coefficients $c_{3,0}$, $c_{3,1}$ and $c_{3,2}$ for the case of Adler $D$-function can be found in Ref.\cite{Dfun}.

To solve this problem, the seBLM method requires the recalculation of the process with the help of multiplets of strongly interacting MSSM gluinos. The purpose of introducing such MSSM gluinos is to determine the $\{\beta_i\}$-coefficients from the $n_f$- and $\tilde{n}_g$- power series; in the final step, one can set $\tilde{n}_g$ to zero in order to obtain the numerical predictions of seBLM for pQCD.

By recalculating the pQCD series with $\tilde{n}_g$ gluinos, the third-order coefficient changes to
\begin{widetext}
\begin{eqnarray}
\tilde{r}_{3}=c_{3,5} \tilde{n}_g n_f +c_{3,4} \tilde{n}_{g}^{2} + c_{3,3} \tilde{n}_g + c_{3,2} n_{f}^{2} + c_{3,1} n_f + c_{3,0}. \label{r3ng}
\end{eqnarray}
\end{widetext}
The coefficients $\beta_0$ and $\beta_1$ with MSSM gluinos are linear functions of $n^{1}_f$ and $\tilde{n}^{1}_g$~\cite{MSSMglu}, thus the $\{\beta_0,\beta_1\}$-coefficients $r_{3}[1]$ and $r_{3}[0,1]$ can be fixed by $c_{3,3}$ and $c_{3,1}$. Note that the $\beta$-series has only four independent coefficients, whereas the $n_f$ and $\tilde{n}_g$ series given above have six expansion coefficients. Thus not all of the $c_{ij}$ coefficients are independent. We can in fact find two relations among $c_{3,5}$, $c_{3,4}$ and $c_{3,2}$; {\it i.e.}
\begin{displaymath}
c^2_{3,5}=4c_{3,2}\cdot c_{3,4}\;\; {\rm and}\;\; 4T^2_R c_{3,4}=C_A^2 c_{3,2},
\end{displaymath}
which agree with the analytic results derived in Ref.\cite{Dfun}. Thus, the $\beta^2_0$-coefficients $r_{3}[2]$ can be fixed by $c_{3,5}$ or $c_{3,4}$ or by $c_{3,2}$.

These seBLM procedures can be extended to all orders. For example, the seBLM $\beta$-pattern at the four-loop level can be written as
\begin{widetext}
\begin{equation}
\beta_0^3\cdot r_{4}[3]+\beta_0 \beta_1\cdot r_{4}[1, 1]+\beta_2\cdot r_{4}[0, 0, 1]+ \beta_0^2\cdot r_{4}[2]+ \beta_1 \cdot r_{4}[0, 1]+ \beta_0 \cdot r_{4}[1]+r_{4}[0]. \label{fourBLMexp}
\end{equation}
\end{widetext}
The pQCD $n_f$-power series at the four-loop level can be written as
\begin{equation}
r_{4}= c_{4,3} n^3_f + c_{4,2} n^2_f + c_{4,1} n_f + c_{4,0}. \label{r4nf}
\end{equation}
Again it cannot fix all the $\{\beta_i\}$-coefficients. By recalculating $r_{4}$ with $\tilde{n}_g$ gluinos, the fourth-order coefficient changes to
\begin{widetext}
\begin{eqnarray}
\tilde{r}_{4}=c_{4,9} \tilde{n}_g n^2_f + c_{4,8} \tilde{n}^2_g n_f + c_{4,7} \tilde{n}_g n_f + c_{4,6} \tilde{n}_{g}^{3} + c_{4,5} \tilde{n}_{g}^{2} + c_{4,4} \tilde{n}_g + c_{4,3} n_{f}^{3} + c_{4,2} n_{f}^{2} + c_{4,1} n_f + c_{4,0}. \label{r4ng}
\end{eqnarray}
\end{widetext}
The $\{\beta_0,\beta_1\}$-coefficients $r_{4}[1]$ and $r_{4}[0,1]$ can be fixed by $c_{4,4}$ and $c_{4,1}$; the $\{\beta_0 \beta_1, \beta_2, \beta_0^2\}$-coefficients $r_{4}[1,1]$, $r_{4}[0,0,1]$ and $r_{4}[2]$ can be fixed by $c_{4,7}$, $c_{4,5}$ and $c_{4,2}$. The $\beta$-series (\ref{fourBLMexp}) has seven independent coefficients, whereas the $n_f$ and $\tilde{n}_g$ series (\ref{r4ng}) have ten expansion coefficients. Thus there are three constraints which constrain the $c_{ij}$ coefficients at this level. Using the equivalence of (\ref{fourBLMexp}) and (\ref{r4ng}), we obtain
\begin{eqnarray}
9c_{4,3}\cdot c_{4,6} &=& c_{4,8}\cdot c_{4,9}, \nonumber\\
C_A^3 c_{4,3} &=& 8T_R^3 c_{4,6}, \nonumber \\
2T_R c_{4,8} &=& C_A c_{4,9}. \nonumber
\end{eqnarray}
Thus, only one of $c_{4,9}$, $c_{4,8}$, $c_{4,6}$ and $c_{4,3}$ is independent; it can be used to fix the $\beta^3_0$-coefficient $r_{4}[3]$. At present, the fourth-order coefficient $\tilde{r}_4$ is not available for any observable; the general relations given above can be used as a cross-check on $\tilde{r}_4$ when it is calculated in the future and as a self-consistency check on the seBLM idea.

The $\tilde{n}_g$-related coefficients are process-dependent. One requires the $\tilde{n}_g$-dependent calculation in order to fix all the $\{\beta_i\}$-coefficients. This unavoidably introduces extra loop calculations into the already complex higher-order QCD corrections. Due to the present limited knowledge of the $\tilde{n}_g$-dependent $\{\beta_i\}$-expression and the $\tilde{n}_g$-dependent pQCD series, the seBLM can only be applied to deal with Adler $D$-function-derived processes, which is now only known at the three-loop level. For future usage, we present the seBLM coefficients $r_{k}[l, m, \cdots]$ at the scale $\mu=Q$ for the Adler $D$-function in the Appendix \ref{R coefficient}, where $Q$ stands for the measured physical scale.

\subsubsection{Determination of the seBLM scales}

After fixing the $\beta$-pattern and the $\{\beta_i\}$-coefficients for a process, the ``large $\beta_0$-approximation" can be adopted with a slight alteration to analyze the pQCD series. This first step can be done at any order. As an explanation and for the subsequent use of MseBLM, we will present a four-loop analysis of the pQCD series. Up to four-loop level, the pQCD approximant $\varrho_n$ can be formally written as
\begin{widetext}
\begin{eqnarray}
\varrho_n=&&r_{0} +r_{1} \left[ a(\mu)+\left(\underline{\beta_0 \cdot r_{2}[1]}+ r_{2}[0]\right )a^{2}(\mu)+\left(\underline{\beta_0^2\cdot r_{3}[2]+\beta_1\cdot r_{3}[0,1]}+ \beta_0\cdot r_{3}[1] + r_{3}[0]\right)a^{3}(\mu) \right. \nonumber\\
&& \left. +\left(\underline{\beta_0^3\cdot r_{4}[3]+\beta_0 \beta_1\cdot r_{4}[1,1]+\beta_2\cdot r_{4}[0,0,1]}+ \beta_0^2 \cdot r_{4}[2]+ \beta_1 \cdot r_{4}[0,1]+ \beta_0 \cdot r_{4}[1]+r_{4}[0]\right )a^{4}(\mu) + \cdots \right], \label{seBLM0}
\end{eqnarray}
\end{widetext}
We note that in Ref.\cite{Kataev1}, the elimination of $\{\beta_i\}$-terms starts with the so-called RG-invariant $\varrho_n|_{\mu=Q}$. This is not a strict condition since the fixed-order pQCD approximate $\varrho_n|_{\mu=Q}$ obtained using conventional scale setting cannot be a RG-invariant. Here, we keep $\mu$ arbitrary; its value only needs to be large enough to ensure a pQCD calculation. The key idea of seBLM is to use the relation $\beta_i \sim \beta^{i+1}_0$ in order to rearrange all the terms at the same order following the equivalent $\beta_0$-powers, and then to eliminate the $\{\beta_i\}$-terms sequentially.

The first step is to set the scale $Q_1$ at the NLO level, which is determined by eliminating all the $\{\beta_i\}$-terms with highest equivalent $\beta_0$-power at each perturbative order; i.e. by absorbing all the underlined terms of Eq.(\ref{seBLM0}) into the running coupling:
\begin{widetext}
\begin{equation}
\varrho'_n=r_{0} + r_{1} a(Q_1) \left(1+r_{2}[0] a(Q_1)+(\underline{\underline{\beta_0\cdot \widetilde{r_{3}[1]}}}+ r_{3}[0])a^{2}(Q_1) +(\underline{\underline{\beta_0^2\cdot \widetilde{r_{4}[2]}+ \beta_1 \cdot \widetilde{r_{4}[0,1]}}}+ \beta_0 \cdot \widetilde{r_{4}[1]}+r_{4}[0])a^{3}(Q_1)+ \cdots \right), \label{seBLM1}
\end{equation}
\end{widetext}
where the {\it tilde} symbol means its value differs from the untilded one. The seBLM scale $Q_1$ satisfies
\begin{equation}
\ln \frac{\mu^{2}}{Q_{1}^{2}} = \Delta_{1,0}+\Delta_{1,1} (\beta_{0}\cdot a(Q_1))+\Delta_{1,2} (\beta_{0}\cdot a(Q_1))^{2} + \cdots, \label{scaleQ1}
\end{equation}
where $\Delta_{1,0}$, $\Delta_{1,1}$ and $\Delta_{1,2}$ are used to eliminate the underlined terms of Eq.(\ref{seBLM0}) at the N$^2$LO, N$^3$LO and N$^4$LO levels, respectively.

The second step is to set the scale $Q_2$ at the N$^2$LO level, which is determined by absorbing the doubly-underlined terms into the running coupling
\begin{widetext}
\begin{equation}
\varrho''_n=r_{0}+r_{1}a(Q_1) \left(1+a(Q_2)\left(r_{2}[0]+r_{3}[0] a(Q_2)+(\underline{\underline{\underline{\beta_0 \cdot \widetilde{\widetilde{r_{4}[1]}}}}}+r_{4}[0])a^{2}(Q_2)+\cdots\right)\right). \label{seBLM2}
\end{equation}
\end{widetext}
The seBLM scale $Q_2$ satisfies
\begin{equation}
\ln \frac{Q_{1}^{2}}{Q_{2}^{2}} = \Delta_{2,0}+\Delta_{2,1} (\beta_{0} a(Q_2))+\cdots, \label{scaleQ2}
\end{equation}
where $\Delta_{2,0}$ and $\Delta_{2,1}$ are used to eliminate the doubly-underlined terms of Eq.(\ref{seBLM1}) at the N$^3$LO and N$^4$LO levels, respectively.

The third step is to set the scale $Q_3$ at the N$^3$LO level, which is derived by absorbing the triply-underlined terms of Eq.(\ref{seBLM2}) into the running coupling
\begin{widetext}
\begin{equation}
\varrho'''_n=r_{0}+r_{1} a(Q_1) \left(1+a(Q_2)\left(r_{2}[0]+r_{3}[0] a(Q_3)+r_{4}[0] a^{2}(Q_3) +\cdots\right)\right) . \label{seBLM3}
\end{equation}
\end{widetext}
Eq.(\ref{seBLM3}) is the final seBLM predictions for the pQCD approximant $\varrho_n$, in which the seBLM scale $Q_3$ satisfies
\begin{equation}
\ln \frac{Q_{2}^{2}}{Q_{3}^{2}} = \Delta_{3,0}+\cdots, \label{scaleQ3},
\end{equation}
where $\Delta_{3,0}$ is used to eliminate the triply-underlined terms of Eq.(\ref{seBLM2}) at the N$^4$LO level. As with the PMC, residual scale dependence remains due to unknown high-order $\{\beta_i\}$-terms.

For definiteness, we will also adopt $Q_4=Q_3$ to perform the seBLM predictions. The expressions for all the $\Delta_{i,j}$ coefficients can be found in Appendix \ref{delta expansion}. In the above derivation, we have adopted the equivalent $\beta_0$-powers~\cite{Kataev1} to deal with the $\{\beta_i\}$-series at each order. We have checked that one can directly replace all $\beta_i$ by $c_i \beta_0^{i+1}$ to do the scale setting, and obtain exactly the same seBLM predictions.

\subsubsection{MseBLM}

As already discussed, the seBLM has some weak points which constrain its applicability. In particular, all $n_f$-terms are eliminated by the seBLM method without distinguishing whether those $n_f$-terms pertain to the RG $\beta$-function. The seBLM procedure has been illustrated by comparing the conformal symmetry limit of QED and QCD for the case of the Adler $D$-function. That is, by taking the Abelian $N_c\to 0$ limit of the $SU(N_c)$ group parameters, such as $C_F=1$, $C_A=0$, $T_F=1$, $f^{abc}=1$ and $d_F^{abc}=1$, together with the condition $n_f=0$, the seBLM gets the same perturbative series for the $D$-function as that of ``quenched QED" (QED with $n_f=0$). Since the quenched QED leads to a conformal series~\cite{Baikov1}, it is thus concluded that the seBLM pQCD series is also conformal in this case~\cite{Kataev5}. However, this argument, being based on the Abelian limit, is not valid in general; in particular, it is invalid for high-order pQCD predictions, because by taking the Abelian limit of the QCD series, all the three-gluon and four-gluon couplings are absent~\cite{Baikov1} many of which also contain conformal contributions.

Let us discuss these issues based on the RG point-of-view: RG-invariance states that a physical quantity must be independent of the renormalization scale and scheme~\cite{RGI1,RGI2,RGI3,RGI4}. In general, an anomalous dimension must also be introduced to ensure the RG-invariance~\cite{beta0,beta00,beta01,beta02,MSbar}, such as in the case of the Adler $D$-function. As mentioned in the introduction, RG-invariance is broken at fixed order, leading to well-known residual renormalization scale- and scheme-ambiguities. If one requires a {\it fixed-order} prediction to satisfy RG-invariance, as suggested by Stevenson~\cite{PMS1,PMS2,PMS3} (called local RG-invariance~\cite{pmc6}), one can derive an ``optimal scale" and even an ``optimal scheme" of a process by using the extended RG-equations. This is the method of the ``Principal of Minimum Sensitivity (PMS). However, since the standard (global) RG-invariance is broken, the PMS predictions do not satisfy basic RG properties~\cite{pmccolloquium}, and its pQCD convergence is accidental. In addition, it fails to achieve the correct prediction of higher-order contributions when one only knows the NLO correction~\cite{PMS4}. This limitation also explains why the predicted PMS scale for the NLO three-jet production via $e^+ e^-$-annihilation does not  yield the correct physical behavior for low $e^+ e^-$ collision energy~\cite{jet1,jet2}. A detailed comparison of the PMS and the PMC can be found in Refs.~\cite{PMS4}.

We find that not distinguishing the $n_f$-terms is in some sense equivalent to using local RG-invariance to set the renormalization scales. More explicitly, Eqs.(2.12-2.16,2.22) in Ref.~\cite{Kataev5} agrees with the PMS scale equation. Thus seBLM will in principle meet the same problems of PMS. By eliminating the $n_f$-terms in the anomalous dimension function simultaneously with the $n_f$-terms for renormalizing the running coupling, the seBLM may achieve effective scale of the process for improving pQCD convergence, but it cannot determine the correct behavior of the running coupling.

On the other hand, the $\beta$-pattern and correct $\{\beta_i\}$-coefficients using the RG-equation and the standard RG-invariance are correctly determined using the PMC. A central goal of seBLM is to improve the pQCD convergence; we can retain this goal by a modification which will improve its applicability; {\it i.e.,} we can use the PMC method to determine the $\{\beta_i\}$-coefficients to replace the seBLM method, while keeping the seBLM procedures for eliminating the $\beta$-terms. We will call this modified method MseBLM. The MseBLM inherits the main seBLM properties, but it avoids the introduction of extra MSSM gluinos, thus making it applicable to any process and to any order.

In distinction to seBLM, the MseBLM takes the same $\beta$-pattern and the same $\{\beta_i\}$-coefficients as those of PMC. It should however only absorb those $n_f$-terms which correctly determine the behavior into the coupling constant; thus the coefficients for the $\{\beta_i\}$-terms are different from the seBLM ones. We shall show that the correct $\{\beta_i\}$-terms are not only needed for achieving the optimal running coupling, but they are also important for improving pQCD convergence. This partly explains why the seBLM and PMC predictions listed in Ref.\cite{Kataev1} behave quite differently \footnote{Note that due to the misunderstanding of the PMC procedures, the PMC predictions in Ref.\cite{Kataev1} are incorrect.}.

\subsection{Formulas for the Adler D-Function}

It should be emphasized that the anomalous dimension function which appears in the Adler $D$-function has no relation to the $\alpha_s$-renormalization of the process; it needs to be separately kept fixed during
PMC scale-setting in order to obtain the correct $\alpha_s$-running behavior~\cite{BMW2}. The $\{\beta_i\}$-terms that should be absorbed into the running coupling in the Adler $D$-function can be written as
\begin{widetext}
\begin{eqnarray}
D(a) &=& 12\pi^2\left(\gamma(a)-\beta(a) \frac{d }{d a}\Pi(a)\right) = \left(\sum_f q_{f}^{2}\right) d_R D^{\rm NS}(a)+ \left(\sum_f q_{f}\right)^{2} D^{\rm S}(a)  \label{D} \\
D^{\rm NS}(a) &=& 1+ 4a + \left(12\gamma_{2}^{\rm NS}+3\beta_0 \Pi_{1}^{\rm NS}\right) a^2 + \left(48\gamma_{3}^{\rm NS} +3\beta_1 \Pi_{1}^{\rm NS}+24\beta_0 \Pi_{2}^{\rm NS}\right) a^3 \nonumber\\
&& +\left(192\gamma_{4}^{\rm NS} +3\beta_2 \Pi_{1}^{\rm NS}+24\beta_1 \Pi_{2}^{\rm NS} +144\beta_0 \Pi_{3}^{\rm NS}\right) a^4 \label{D function} \\
D^{\rm S}(a) &=& 48\gamma_{3}^{\rm S} a^3+ \left(192\gamma_{4}^{\rm S}+144\beta_0 \Pi_{3}^{\rm S}\right) a^4 + \cdots ,
\end{eqnarray}
\end{widetext}
where $q_f$ stands for the electric charge of the active flavors. The subscript ``NS" and ``S" denote the non-singlet and singlet parts. $d_R=N_c$ in the fundamental representation of the $SU(N_c)$ group. $\gamma_{i}^{\rm (N)S}$ and $\Pi_{j}^{\rm (N)S}$ can be found in Ref.\cite{Baikov}. There are no $\beta_0^2$-terms at the three-loop level, but they are present in the anomalous dimension $\gamma_{3}^{\rm NS}$ and $\gamma_{3}^{\rm S}$ and are unrelated to $\alpha_s$-renormalization~\footnote{For the anomalous dimension itself, one may need to apply the PMC to achieve a better pQCD prediction, which however is out of the scope of the present paper.}. The $\beta$-pattern is different from the seBLM one~\cite{Kataev1}, in which the $\{\beta_i\}$-terms from the anomalous dimensions are incorrectly included in order to obtain the $\beta$-pattern. In the following, we shall show that the correct $\beta$-pattern together with correct $\{\beta_i\}$-term coefficients are essential for the correct pQCD prediction.

\subsection{Formulas for $R_{e^+ e^-}$}

The ratio of $e^+e^-$ annihilation into hadron over muon-pairs $R_{e^+ e^-}$ provides one of the most precise tests of pQCD. The measured observable $R(Q)$ is defined as:
\begin{equation}
R_{e^+e^-}(Q) = \frac{\sigma (e^+e^-\to {\rm hadrons})}{\sigma (e^+e^-\rightarrow \mu^+\mu^-)} =3 \sum_{f}q_{f}^{2} (1+R(Q)),
\end{equation}
where $Q= \sqrt{s} = E_{CM}$ stands for the $e^+e^-$-collision energy at which it is measured. The timelike $R_{e+e-}$ is related to the Adler $D$-function through the equation
\begin{equation}
R_{e+e-}(s)=\frac{1}{2\pi i} \int_{-s-i \epsilon}^{-s+i \epsilon} \frac{D(q^2)}{q^2} d q^2.
\end{equation}
The pQCD approximation for the $R$-ratio $R(Q)$ is
\begin{equation}
R_n(Q,\mu) = \sum^{n}_{i=1} {\cal C}_i(Q,\mu) a^{i}(\mu),
\end{equation}
where $a=\alpha_s/4\pi$. Using the Adler $D$-function, we can obtain the four-loop pQCD approximant $R_4(Q)$, with the initial coefficients of the explicit PMC or MseBLM $\beta$-pattern at the scale $\mu=Q$ given by~\cite{Kuhn}
\begin{eqnarray}
{\cal C}_{1} &=& 3\gamma_{1}^{\rm NS}=4, \\
{\cal C}_{2} &=& 12\gamma_{2}^{\rm NS} +3\beta_0 \Pi_{1}^{\rm NS}, \\
{\cal C}_{3} &=& 48\left(\gamma_{3}^{\rm NS}+ \frac{\left(\sum_f q_{f}\right)^{2}} {3 \sum_f q_{f}^{2}}\gamma_{3}^{\rm S}\right)+24\beta_0 \Pi_{2}^{\rm NS} \nonumber\\
&& +3\beta_1 \Pi_{1}^{\rm NS}-(\pi \beta_{0})^{2}\gamma_{1}^{\rm NS}
\end{eqnarray}
and
\begin{eqnarray}
{\cal C}_{4} &=& 192\left(\gamma_{4}^{\rm NS}+\frac{\left(\sum_f q_{f}\right)^{2}}{3 \sum_f q_{f}^{2}} \gamma_{4}^{\rm S}\right) +3\beta_2 \Pi_{1}^{\rm NS} \nonumber\\
&& +24\beta_1 \Pi_{2}^{\rm NS}+144\left(\Pi_{3}^{\rm NS}+\frac{\left(\sum_f q_{f}\right)^{2}}{3 \sum_f q_{f}^{2}}\Pi_{3}^{\rm S}\right) \beta_0 \nonumber\\
&&-12(\pi \beta_{0})^{2}\gamma^{\rm NS}_2 -\frac{5}{2}\pi^2 \beta_{0} \beta_1 \gamma^{\rm NS}_1 -3\pi^2\beta_0^3\Pi^{\rm NS}_1.
\end{eqnarray}
The coefficients ${\cal C}_i$ at other scales can be obtained via the scale displacement relation (\ref{scaledis}).

Following the standard procedures for seBLM, MseBLM and PMC, one can achieve the predictions under various scale-setting approaches. The resulting pQCD approximation for the $R$-ratio is
\begin{equation}
R_n(Q,\mu) = \sum^{n}_{i=1} \tilde{\cal C}^{\rm SS}_i(Q) a^{i}(Q^{\rm SS}_i(Q,\mu)),
\end{equation}
where $\tilde{\cal C}^{\rm SS}_i(Q)$ are free of $\{\beta_i\}$-terms and free of initial scale choice. Here the $Q^{\rm SS}_i$ are the effective scales for each scale-setting approach where ${\rm SS}$ stands for seBLM, MseBLM and PMC, respectively.

\section{Numerical results for $R_{e^+ e^-}$} \label{analysis}

In order to obtain numerical predictions for $R_n$, the QCD parameters will be fixed using $\alpha_s(M_Z)=0.1185\pm0.0006$~\cite{pdg}. To be consistent, we shall adopt the $n_{\rm th}$-loop $\alpha_s$-running to do the calculation, and the $\Lambda_{\overline{\rm{MS}}}$ is determined by using the $n_{\rm th}$-loop $\alpha_s$-running determined from the RG-equation. For example, we obtain $\Lambda_{\overline{\rm{MS}}}^{(n_f=5)} =214$ MeV for $R_4$ by using the four-loop $\alpha_s$-running. It is found that the residual scale dependence for all scale settings are highly suppressed; if not specially stated, we will take the initial scale $\mu\equiv Q$. For definiteness, we will set $Q=31.6$ GeV~\cite{experiment} to do our analysis.

\subsection{A comparison of three-loop $R_{3}$}

\begin{table}[htb]
\begin{tabular}{ccccc}
\hline
              & ~~$\tilde{\cal C}_1$~~ & ~~$\tilde{\cal C}_2$~~   & ~~$\tilde{\cal C}_3$~~    \\
\hline
Conv.     & 4   & 22.548  & -819.496   \\
PMC      & 4   & 29.444  & -64.248  \\
seBLM     & 4   & 1.333  & -$2.298\times10^3$   \\
\hline
\end{tabular}
\caption{Final expansion coefficients $\tilde{\cal C}_n$ for $R_{3}$ after applying conventional (Conv.), PMC and seBLM scale settings. The T=three-loop $\alpha_s$-running is adopted. $Q=31.6$ GeV. } \label{Reecoe}
\end{table}

The seBLM stops at the N$^2$LO QCD corrections. After applying PMC or seBLM, we can fix two PMC or seBLM scales from the known $\{\beta_i\}$-terms, which are
\begin{eqnarray}
Q^{\rm PMC}_{1}&=&40.84\;{\rm GeV},\; Q^{\rm PMC}_{2}=29.79\;{\rm GeV}, \\
Q^{\rm seBLM}_{1}&=&24.38\;{\rm GeV},\; Q^{\rm seBLM}_{2}=2.52 \times 10^{-35}\;{\rm GeV}.
\end{eqnarray}
It is seen that the PMC scales are physically reasonably, whereas the seBLM scale $Q^{\rm seBLM}_2$ is far from the pQCD domain, making the final pQCD prediction unreliable. This was already observed by Refs.\cite{seBLM1,Kataev1}, which also indicates the importance of the correct determination of $\{\beta_i\}$-terms. We present the final coefficients $\tilde{\cal C}_n$ with $n=(1,2,3)$ after applying conventional (Conv.), PMC and seBLM scale settings in Table \ref{Reecoe}. One finds $\tilde{\cal C}_n\equiv {\cal C}_n$ for conventional scale setting. After applying PMC, the coefficients $\tilde{\cal C}_n$ are the scheme-independent conformal terms. After applying the seBLM, the coefficients $\tilde{\cal C}_n$ are the remaining terms by eliminating all $n_f$-terms. As shown by Table \ref{Reecoe}, after applying the PMC, the third-order coefficient $\tilde{\cal C}_3$ is much smaller than the conventional one; in contrast, after applying the seBLM, the third-order coefficient $\tilde{\cal C}_3$ is even larger than the conventional one. We point out that a comparison of the PMC and seBLM coefficients and scales after finishing only the first step of the procedures, as was done in Ref.\cite{Kataev1}, is inconsistent.

\begin{table}[htb]
\begin{tabular}{cccccc}
\hline
        & LO   & NLO   & N$^2$LO    & $total$ \\
\hline
Conv.     &0.04497	 & 0.00285	&-0.00116	&0.04666 \\
PMC      &0.04296	 & 0.00380 &-0.00009	&0.04667 \\
seBLM     &0.04721	 & -    & -	  &- \\
$x$BLM     &0.04622 & 0.00018 & 0     &0.04640 \\
\hline
\end{tabular}
\caption{Each QCD loop's contribution to $R_{3}$ under conventional (Conv.), PMC and seBLM scale settings. The $total$-column stands for the sum of all those loop corrections. The $x$BLM results are also presented. Three-loop $\alpha_s$-running is adopted. $Q=31.6$ GeV.} \label{Ree3loop}
\end{table}

We present the separate contributions from the one-loop (LO), two-loop (NLO) or three-loop (N$^2$LO) QCD contribution to $R_{3}$ under PMC, seBLM and conventional scale settings in Table \ref{Ree3loop}. Due to the elimination of divergent renormalon terms, the PMC pQCD convergence is better than that of conventional scale setting, i.e. $|R_{3,\rm PMC}^{\rm LO}| \gg |R_{3,\rm PMC}^{\rm NLO}| \gg |R_{3,\rm PMC}^{\rm N^2LO}|$.

The seBLM is designed to improve the pQCD convergence by applying the ``large $\beta_0$-approximation". However, the seBLM pQCD convergence for $R_{e^+ e^-}$ becomes even worse than the conventional one, and -- since it involves an unreasonable small scale $Q^{\rm seBLM}_2$ -- the seBLM pQCD prediction is questionable. To cure this problem, a seBLM alteration, i.e. the $x$-BLM~\cite{seBLM1} or equivalently the one-scale seBLM~\cite{Kataev1}, has been suggested. Such an alteration requires an overall modification of the total pQCD prediction by directly requiring the N$^2$LO-term $\tilde{\cal C}_3$ to be zero; i.e. $\tilde{\cal C}^{x{\rm BLM}}_1=4$, $\tilde{\cal C}^{x{\rm BLM}}_2=1.333$ and $\tilde{\cal C}^{x{\rm BLM}}_3=0$. This treatment makes the pQCD prediction more reliable as shown by Table \ref{Ree3loop}. However, it breaks the expected pQCD convergence since the high-order terms $\tilde{\cal C}^{x{\rm BLM}}_{i\geq4}$ are in general nonzero. Moreover, it cannot improve the seBLM applicability due to the introducing of auxiliary fields, which also stops at the N$^2$LO level.

In the following, we shall adopt MseBLM, as an alteration of seBLM, to do a four-loop estimation.

\subsection{A comparison of four-loop $R_{4}$}

\begin{table}[htb]
\begin{tabular}{ccccccc}
\hline
  & $R_2$ & $R_3$  & $R_4$  & $\kappa_1$ & $\kappa_2$ & $\kappa_3$ \\
  \hline
Conv.    &0.04785 &0.04666	&0.04635	&7.4\%	 &-2.5\% &-0.7\% \\
PMC     &0.04767 &0.04667 &0.04637  &7.0\%  &-2.1\% &-0.6\% \\
MseBLM   &0.04767 &0.04654	&0.04640	&7.0\%	 &-2.4\% &-0.3\% \\
 \hline
\end{tabular}
\caption{Numerical results for $R_n$ and $\kappa_n$ up to four-loop level under conventional scale setting (Conv.), PMC and MseBLM. The value of $R_1$ =0.04454 is the same for all scale settings. $Q=31.6$ GeV. } \label{Reekappa}
\end{table}

We first present an overview of how QCD loop corrections affect the pQCD estimates. Numerical results for $R_n$ ($n\leq4$) under conventional scale setting (Conv.), PMC and MseBLM are presented in Table \ref{Reekappa}. To show how the theoretical prediction changes as more-and-more loop corrections are included, we define a ratio:
\begin{equation}
\kappa_n=\frac{R_{n+1}-R_{n}}{R_{n}},  \label{kappa}
\end{equation}
where $n = (1,2,3)$. This ratio indicates how the `known' lower-order estimate is affected by a `newly' available higher-order correction. At the one-loop level, we have no information to set the scale for $R_1$, so we take its scale as $Q$ and we obtain $R_1(Q)=0.04454$ for all scale settings. Table \ref{Reekappa} shows that one can achieve acceptable pQCD predictions with increasing loop corrections from all scale settings. The $\kappa_n$ values for all scale settings behave very similarly. The ratio $|\kappa_3|$ for each scale setting is smaller than $1\%$ up to the four-loop level, indicating that the four-loop pQCD predictions for $R(Q)$ are sufficiently precise.

\begin{table}[htb]
\begin{tabular}{cccccc}
\hline
      & LO   & NLO   & N$^2$LO  & N$^3$LO  & $total$  \\
\hline
Conv.    &0.04499 &0.00285	&-0.00117	&-0.00033	 &0.04634 \\
PMC     &0.04290 &0.00352 &-0.00004	&-0.00002	 &0.04636 \\
MseBLM   &0.04294 &0.00352	&-0.00004	&-0.00001	 &0.04641 \\
\hline
\end{tabular}
\caption{The LO, NLO, N$^2$LO and N$^3$LO loop contributions for the approximant $R_4$ assuming conventional scale setting (Conv.), PMC, seBLM and MseBLM. The $total$-column stands for the sum of all of those loop corrections. Four-loop $\alpha_s$-running is adopted. $Q=31.6$ GeV. } \label{Ree4loop}
\end{table}

\begin{table}[htb]
\begin{tabular}{ccccc}
\hline
   & ~~$\tilde{\cal C}_1$~~ & ~~$\tilde{\cal C}_2$~~ & ~~$\tilde{\cal C}_3$~~ & ~~$\tilde{\cal C}_4$~~ \\
\hline
PMC      & 4   & 29.444  & -64.248  & -$2.813\times10^3$ \\
MseBLM     & 4   & 29.444  & -64.248  & -$2.813\times10^3$ \\
Conv.     & 4   & 22.548  & -819.496 & -$2.059\times10^4$ \\
\hline
\end{tabular}
\caption{Final expansion coefficients $\tilde{\cal C}_n$ for $R_{4}$ after applying the PMC and MseBLM. The expansion coefficients for conventional scale setting are also presented as a comparison. Four-loop $\alpha_s$-running is adopted. $Q=31.6$ GeV.} \label{Reecoe1}
\end{table}

Next, we present a comparison of pQCD convergence assuming various scale settings in Table \ref{Ree4loop}. The standard pQCD convergence under conventional scale setting is due to the $\alpha_s$-power suppression alone. The PMC and MseBLM pQCD series follow the pattern of the standard pQCD series, but are more convergent. As shown in Table \ref{Reecoe1}, after applying PMC and MseBLM, the divergent renormalon terms are eliminated and we get much smaller expansion coefficients $\tilde{\cal C}_n$ at higher orders. The MseBLM and PMC have the same coefficients $\tilde{\cal C}_n$. This explains why a more convergent pQCD series can be achieved for both PMC and MseBLM.

\begin{table}[htb]
\begin{tabular}{ccccccc}
\hline
        & $Q_{1}$ &$Q_{2}$ & $Q_{3}$   \\
\hline
PMC       &~41.23 GeV~ &~36.91 GeV~ &~171.43 GeV~ \\
MseBLM     &41.03 GeV &33.61 GeV & $7.23$ TeV \\
\hline
\end{tabular}
\caption{The determined PMC and MseBLM scales for $R_{4}$. Four-loop $\alpha_s$-running is adopted. $Q=31.6$ GeV. } \label{Reescale3}
\end{table}

It is important to find the correct $\beta$-pattern of a process. The PMC respects RG-invariance and improves the perturbative series by absorbing all $\beta$-terms governed by the RG-equation into the running coupling. Thus, the PMC eliminates the factorial growth of the renormalon nonconformal series and determines and thus yields more precise pQCD predictions. The MseBLM adopts the same $\beta$-pattern as that of PMC, and as shown by Tables \ref{Reekappa} and \ref{Ree4loop}, its predictions are close to those of PMC. There are slight differences for the pQCD series due to different way of absorbing the $\{\beta_i\}$-terms into the running coupling; {\it i.e.} in contrast to the PMC, the MseBLM absorbs the $\{\beta_i\}$-terms via the ``large $\beta_0$-approximation".

We have presented the PMC and MseBLM scales for $R_4$ in Table \ref{Reescale3}; one finds different effective running couplings at each order. Following the procedures in Sec. II, the PMC and MseBLM scales are themselves given as perturbative series. The differences between the PMC and MseBLM scales are suppressed by the accuracy of the approximation $\beta_i \sim \beta^{i+1}_0$ and $\alpha_s$-suppression. The scale differences are formally suppressed by the inverse of equivalent $\beta_0$-powers. In the case of the PMC and MseBLM scales with the highest equivalent $\beta_0$-power, we find a smaller scale difference. This qualitatively explains why the PMC and MseBLM scale differences become larger at higher-orders. In the case of the LO scale $Q_1$, it has the highest equivalent $\beta_0$-power and thus the scale difference is the smallest. More explicitly, we have
\begin{equation}
\Delta Q_1 < \Delta Q_2 < \Delta Q_3,
\end{equation}
where $\Delta Q_i=|Q^{\rm PMC}_{i} - Q^{\rm MseBLM}_{i}|$.

\begin{figure}[htb]
\includegraphics[width=0.50\textwidth]{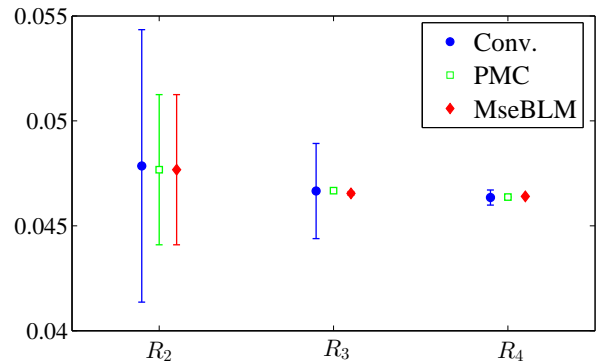}
\caption{Results for $R_n$ ($n=2,3,4$) together with their errors $\left(\pm |\tilde{\cal C}_{n} a^{n+1}|_{\rm MAX} \right)$ at $Q=31.6$ GeV. The big dots, the squares and the diamonds are for conventional scale setting (Conv.), PMC and MseBLM, respectively.} \label{error}
\end{figure}

Finally, we compare the predictive power of PMC and MseBLM. We adopt the conservative procedure suggested by Ref.\cite{pmc6} to predict the ``unknown" high-order pQCD corrections; {\it i.e.,} we identify the perturbative uncertainty with the last known order. Thus the ``unknown" high-order pQCD correction is taken as $\left(\pm |\tilde{\cal C}_{n} a^{n+1}|_{\rm MAX} \right)$ for $R_n$, where $|\tilde{\cal C}_{n} a^{n+1}|$ is calculated by varying $\mu\in[Q/2,2Q]$, and the symbol ``MAX'' stands for the maximum $|\tilde{\cal C}_{n} a^{n+1}|$ within this scale region. The error estimates for conventional, PMC and MseBLM scale settings are displayed in FIG.(\ref{error}). It shows that both the PMC and MseBLM errors are smaller than those assuming conventional scale setting. The PMC and MseBLM errors tend to shrink much more rapidly with the increment of pQCD order, consistent with the pQCD convergence shown by Table \ref{Ree4loop}.

\section{Summary}

In this paper, we have presented a detailed comparison of PMC and seBLM scale settings. Both of these methods are BLM-like approaches with the purpose of extending BLM up to all orders.

The PMC provides the underlying principle for BLM, which sets the optimal scale for any QCD processes up to all orders via a systematic and process-independent way. Following the superposition of the RG-displacement equation, the PMC also provides a direct explanation for the seBLM $\beta$-pattern. The $\{\beta_i\}$-terms are then eliminated by the PMC by absorbing them into the renormalization scales consistent with the RG equation.

Predictions for physical observables should be independent of the choice of the schemes or theoretical conventions. The PMC respects RG-invariance, and the final expression is scheme and scale independent at each finite order. There is a small residual scale dependence due to unknown higher-order $\{\beta_i\}$-terms which is highly suppressed even for low-order predictions. The PMC improves pQCD convergence due to elimination of divergent renormalon terms. The PMC can also be applied to processes with multiple physical scales; e.g., $\Upsilon(1S)$ leptonic decays~\cite{app5}. The effective number of flavors $n_f$ is set correctly at each order of perturbation theory. The PMC is consistent with the standard scale-setting procedure of Gell Mann and Low in the Abelian limit.

The seBLM is designed to improve the pQCD convergence, but it cannot be regarded as a solution for solving renormalization scheme-and-scale ambiguities. All of the seBLM scales are consistent with a $(\beta_0\cdot a)$-power series, and the ``large $\beta_0$-approximation". It sets the $\beta$-pattern approximately via the relation, $\beta_i\sim\beta^{i+1}_0$, and the $\{\beta_i\}$-coefficients are fixed by introducing extra MSSM gluinos. The predictions of the seBLM method strongly depend on the knowledge of $\tilde{n}_g$-dependent pQCD series and $\tilde{n}_g$-dependent $\beta$-functions. At present, the seBLM is only applicable to Adler $D$-function involved processes, and it can only be applied to the three-loop level. However, In contradiction to its main goal, we find that the pQCD convergence of seBLM is questionable, as shown by its application to $R_{e^+ e^-}$.

In order to cure the seBLM shortcomings, we have suggested a modification,
called MseBLM, by borrowing the PMC idea to set the $\{\beta_i\}$-coefficients, while keeping the ``large $\beta_0$ approximation" to deal with the $\beta$-series. It does not require the introduction of extra MSSM gluinos into pQCD calculations. The MseBLM inherits the seBLM properties and makes it applicable to any order.
By taking the four-loop calculation of $R_{e^+ e^-}$ as an example, we have shown that the MseBLM and PMC predictions are numerically consistent; thus more precise pQCD approximation and a more convergent pQCD series for $R_{e^+ e^-}$ can be achieved for both the PMC and the MseBLM. This emphasizes the importance of the correct knowledge of the $\beta$-pattern and the $\{\beta_i\}$-coefficients.

In conclusion, the modified seBLM -- the MseBLM -- provides a practical approach for improving the convergence of pQCD predictions. When more QCD loop terms are considered, it can achieve precise predictions consistent with those of the PMC. However, the PMC has a rigorous theoretical foundation, satisfying all self-consistency conditions from RG-invariance. It thus eliminates an unnecessary systematic error for high precision pQCD predictions, and it can be widely applied to high-energy processes. \\

\noindent{\bf Acknowledge:} This work was supported in part by the Natural Science Foundation of China under Grant No.11275280, the Fundamental Research Funds for the Central Universities under Grant No.CQDXWL-2012-Z002, and the Department of Energy Contract No.DE-AC02-76SF00515. SLAC-PUB-16236. \\

\appendix

\section{The $\{\beta_i\}$-coefficients for the Adler $D$-function}  \label{R coefficient}

Following the standard seBLM procedures, we obtain the $\{\beta_i\}$-coefficients $r_{k}[l, m, \cdots]$ for the Adler $D$-function up to three-loop level. At the scale $\mu=Q$, we have
\begin{eqnarray}
r_1 &=& 3C_F=4 \\
r_2[1] &=& \frac{11}2-4\zeta_3\approx 0.691772 \\
r_2[0] &=& \frac{C_A}3-\frac{C_F}2=\frac{1}3 \\
r_3[2] &=& \frac{302}9-\frac{76}3\zeta_3\approx 3.10345 \\
r_3[0,1] &=& \frac{101}{12}-8\zeta_3\approx -1.19979 \\
r_3[1] &=& C_A\left(-\frac{3}4 + \frac{80}3\zeta_3 -\frac{40}3\zeta_5\right) \nonumber\\
 && - C_F\left(18 + 52\zeta_3 - 80\zeta_5\right)\approx 55.7005
\end{eqnarray}
\begin{eqnarray}
r_3[0] &=& \frac{1}{36} \left(523C_A^2 + 852 C_A C_F - 414 C_F^2\right) \nonumber\\
      & &-72 C_A^2 \zeta_3 + \frac{5}{24}\left(\frac{176}{3}-128\zeta_3\right) \frac{\left(\sum_f q_f\right)^2}{3\left(\sum_f q_f^2\right)} \nonumber \\
      &\approx& -573.9607 -19.8326\frac{\left(\sum_f q_f\right)^2}{3\left(\sum_f q_f^2\right)},
\end{eqnarray}
where $C_A=3$ and $C_F=4/3$ for the $SU_{f}(3)$-group. These coefficients can be derived from Ref.\cite{Dfun}. There is a typo for the coefficient $r_3[1]$ in Refs.\cite{seBLM1,Kataev1}; i.e., the term $+3/4$ there should be corrected to $-3/4$. The $\{\beta_i\}$-coefficients $r_{k}[l, m, \cdots]$ at other scales can be obtained from the above values via the scale displacement relation (\ref{scaledis}).

\section{The $\Delta_{i,j}$ expressions up to four-loop level}
\label{delta expansion}

\begin{widetext}
\begin{eqnarray}
\Delta_{1, 0} &=& r_{2}[1] \\
\Delta_{1, 1} &=& -c_1 r_{2}[1]+c_1 r_{3}[0, 1]-r_{2}[1]^2+r_{3}[2] \\
\Delta_{1, 2} &=& \frac{1}{2} (2 c_1^2 r_{2}[1]-2 c_1^2 r_{3}[0, 1]+3 c_1 r_{2}[1]^2-6 c_1 r_{2}[1] r_{3}[0, 1]-2 c_1 r_{3}[2]\nonumber\\
&&+2 c_1 r_{4}[1, 1]-2 {c_2} r_{2}[1]+2 {c_2} r_{4}[0, 0, 1]+4 r_{2}[1]^3-6 r_{2}[1] r_{3}[2]+2 r_{4}[3]). \\
\Delta_{2, 0} &=& \frac{r_{3}[1]-2 r_{2}[0] r_{2}[1]}{r_{2}[0]} \\
\Delta_{2, 1} &=& \frac{2 c_1 r_{2}[0]^2 r_{2}[1]-2 c_1 r_{2}[0]^2 r_{3}[0, 1]-c_1 r_{2}[0] r_{3}[1]+c_1 r_{2}[0]r_{4}[0, 1]+r_{2}[0]^2 r_{2}[1]^2}{r_{2}[0]^2} \nonumber\\
&&+ \frac{-2 r_{2}[0]^2 r_{3}[2]+r_{2}[0] r_{2}[1] r_{3}[1]+r_{2}[0] r_{4}[2]-r_{3}[1]^2}{r_{2}[0]^2}\\
\Delta_{3, 0} &=& \frac{r_{2}[0] r_{2}[1] r_{3}[0]+r_{2}[0] r_{4}[1]-2 r_{3}[0] r_{3}[1]}{r_{2}[0] r_{3}[0]},
\end{eqnarray}
\end{widetext}
where $c_i$ is defined via the relation $\beta_i=c_i \beta_0^{i+1}$.

As an application, we adopt these $\Delta_{i,j}$ coefficients to study the renormalization scale dependence. For convenience, we first separate the $Q$-dependence from the $\beta$-coefficient $r_{k}[l, m, \cdots]$, where $Q$ stands for the typical scale of the process or at which it is measured.
\begin{widetext}
\begin{eqnarray}
r_{1}(\mu) &=& r_{1}, \label{ini1}\\
r_{2}(\mu) &=& r_{2}[0]+\beta_0 \left(r_{2}[1]+ \ln\frac{\mu^{2}}{Q^2}\right), \label{ini2} \\
r_{3}(\mu) &=& r_{3}[0]+\beta_0 \left(r_{3}[1]+ 2 r_{2}[0] \ln\frac{\mu^{2}}{Q^2}\right) +\beta_1 \left(r_{3}[0,1] + \ln\frac{\mu^{2}}{Q^2}\right) +\beta_0^2 \left(r_{3}[2]+ 2r_{2}[1] \ln\frac{\mu^{2}}{Q^2}+ \ln^2\frac{\mu^{2}}{Q^2}\right), \label{ini3} \\
r_{4}(\mu) &=& r_{4}[0]+\beta_0^2 \left(3 {r_{2}[0]} \ln^2\frac{\mu^{2}}{Q^2}+3 r_{3}[1] \ln\frac{\mu^{2}}{Q^2} +r_{4}[2]\right)+\beta_1 \left(r_{4}[0,1] + 2 {r_{2}[0]} \ln\frac{\mu^{2}}{Q^2}\right)+\nonumber\\
&&\frac{1}{2} \beta_0 \beta_1 \left(2 r_{4}[1,1] +\left(4 {r_{2}[1]}+6 r_{3}[0,1]\right) \ln \frac{\mu^{2}}{Q^2}+5 \ln^2\frac{\mu^{2}}{Q^2}\right)+ \beta_0\left(r_{4}[1]+3 r_{3}[0] \ln\frac{\mu^{2}}{Q^2}\right)\nonumber\\
&&+\beta_0^3 \left(r_{4}[3]+3 r_{3}[2] \ln\frac{\mu^{2}}{Q^2}+3 {r_{2}[1]} \ln^2\frac{\mu^{2}}{Q^2}+\ln^3\frac{\mu^{2}}{Q^2}\right)+\beta_2 \left( r_{4}[0,0,1]+ \ln\frac{\mu^{2}}{Q^2}\right). \label{ini4}
\end{eqnarray}
\end{widetext}
We have implicitly taken $r_{k}[l, m, \cdots] =r_{k}[l, m, \cdots]|_{\mu=Q}$. For processes with several typical scales, the condition is more involved but can be done via a similar way.
We can read off the scale-dependent $\{\beta_i\}$-coefficients $r_{k}[l, m, \cdots]$, which are
\begin{eqnarray}
r_{2}[1] &\to& r_{2}[1]+ \ln\frac{\mu^{2}}{Q^2} \\
r_{3}[1] &\to& r_{3}[1]+ 2 r_{2}[0] \ln\frac{\mu^{2}}{Q^2} \\
r_{3}[0,1] &\to& r_{3}[0,1] + \ln\frac{\mu^{2}}{Q^2} \\
r_{3}[2] &\to& r_{3}[2]+2r_{2}[1]\ln\frac{\mu^{2}}{Q^2}+ \ln^2\frac{\mu^{2}}{Q^2} \\
r_{4}[2] &\to& 3 {r_{2}[0]} \ln^2\frac{\mu^{2}}{Q^2}+3 r_{3}[1] \ln\frac{\mu^{2}}{Q^2} +r_{4}[2]\\
r_{4}[0,1] &\to& r_{4}[0,1] + 2 {r_{2}[0]} \ln\frac{\mu^{2}}{Q^2} \\
r_{4}[1,1] &\to& r_{4}[1,1] +\left(2 {r_{2}[1]}+3 r_{3}[0,1]\right) \ln \frac{\mu^{2}}{Q^2}\nonumber\\
&& +\frac{5}{2} \ln^2\frac{\mu^{2}}{Q^2}
\end{eqnarray}
\begin{eqnarray}
r_{4}[1] &\to& r_{4}[1]+3 r_{3}[0] \ln\frac{\mu^{2}}{Q^2} \\
r_{4}[3] &\to& r_{4}[3]+3 r_{3}[2] \ln\frac{\mu^{2}}{Q^2}+3 {r_{2}[1]} \ln^2\frac{\mu^{2}}{Q^2} \nonumber\\
&& +\ln^3\frac{\mu^{2}}{Q^2} \\
r_{4}[0,0,1] &\to& r_{4}[0,0,1]+ \ln\frac{\mu^{2}}{Q^2}
\end{eqnarray}

Substituting them into $\Delta_{i,j}$, we find that, except for $\Delta_{1,0}= r_{2}[1] + \ln\frac{\mu^{2}}{Q^2}$, all other $\Delta_{i,j}$ coefficients are free of $\mu$-dependence. For Eq.(\ref{scaleQ1}), we obtain
\begin{displaymath}
\ln \frac{Q^{2}}{Q_{1}^{2}}=r_{2}[1] +\Delta_{1,1} (\beta_{0}\cdot a(Q_1)) +\Delta_{1,2} (\beta_{0}\cdot a(Q_1))^{2}.
\end{displaymath}
This shows that $Q_1$, and thus all the high-order seBLM scales, as indicated by Eqs.(\ref{scaleQ2},\ref{scaleQ3}), should be independent of the initial choice of scale. This property is ensured by the local RG-invariance mentioned in the body of the text.

\end{document}